# Load Balancing Optimization in LTE/LTE-A Cellular Networks: A Review

Sumita Mishra (Corresponding author), Nidhi Mathur

Electronics and Communication Engineering Department,

Amity school of Engineering and Technology,

Amity University, Lucknow Campus

E-mail: mishra.sumita@gmail.com

**Abstract**

During the past few decades wireless technology has seen a tremendous growth. The recent introduction of high-end mobile devices has further increased subscribers' demand for high bandwidth. Current cellular systems require manual configuration and management of networks, which is now costly, time consuming and error prone due to exponentially increasing rate of mobile users and nodes. This leads to introduction of self organizing capabilities for network management with minimum human involvement. It is expected to permit higher end user Quality of Service (QoS) along with less operational and maintenance cost for telecom service providers. Self organized cellular networks incorporate a collection of functions for automatic configuration, optimization and maintenance of cellular networks. As mobile end users continue to use network resources while moving from a cell boundary to other, traffic load within a cell does not remain constant. Thus Load balancing as a part of self organized network solution, has become one of the most active and emerging fields of research in Cellular Network. It involves transfer of load from overloaded cells to the neighbouring cells with free resources for more balanced load distribution in order to maintain appropriate end-user experience and network performance. In this paper, review of various load balancing techniques currently used in mobile networks is presented, with special emphasis on techniques that are suitable for self optimization feature in future cellular networks.

**Keywords:** 4G, Self-Optimization, Load Balancing, eNB, LTE Advanced

## 1. Introduction

Demand for bandwidth is increasing at a rapid pace due to popular use of Smartphone, tablets and other high end web enabled devices. Mobile users and the resulting data usage are random, time varying, and often unbalanced, this make unequal load scenario for neighbour cells; consequently one cell may be overloaded while, other has much less users and its resources are not fully utilized; overloaded cell faces resource shortage, which affects access of new users and impacts the QoS of active users. Thus load imbalance seriously deteriorates the overall performance of the cellular network due to inefficient resource utilization; the problem is rectified by load balancing techniques. Load balancing involves detecting network load imbalance by periodically exchanging information between neighbouring eNBs to compare the cells load and then, Load balancing optimization feature automatically change the network parameters to manage the unexpected loading conditions caused by sudden rise in data usage.

The rest of this paper is organized as follows. Section II briefly provides the background information. Section III begins with a detailed discussion on Load Balancing schemes in cellular network with special emphasis on the schemes suitable for self organized cellular networks, and finally conclusions are drawn in Section IV.

## 2. Overview of LTE/LTE –Advanced

It is expected that future generation of mobile networks will be based on 4G technology. Currently 4G or Fourth generation of mobile/wireless broadband is a collection of wireless standards, and it can be realized using various competing technologies; one of the prominent technologies is Long Term Evolution (LTE) and LTE-Advanced. LTE and LTE-A supports IP (Internet Protocol) based packet switching communication system with OFDM multi-carrier transmission and other frequency domain schemes which offer high data transfer rates, which are further increased by using Antenna arrays for MIMO (Multiple Input Multiple Output) communication to avoid multi-path propagation losses.

LTE is an all IP based system, thus all the data including voice is sent as an IP packet. It gives high peak data rate due to flexibility of spectrum usage with low latency times. Efficient radio usage and optimization procedures ensure higher capacity per cell and Cost-efficient deployment.

The radio interface of LTE is based on Orthogonal Frequency Division Multiple Access (OFDMA) in the downlink and Single Carrier-Frequency Division Multiple Access (SC-FDMA) in the uplink.

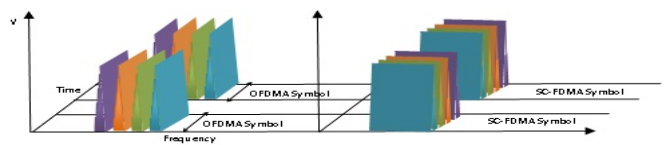

Figure 1: Downlink OFDMA and Uplink SC-FDMA

| Downlink frame (OFDMA) | | | | | |
|---|---|---|---|---|---|
| Preamble | Header | Burst-1 | Burst-2 | Brust-3 | ------------ |
| Uplink frame (SC-FDMA) | | | | | |
| Preamble | Burst-1 | Preamble | Brust-2 | Preamble | Brust-3 | ---------- |

Table 1: Downlink and Uplink frames

LTE uses All IP based flat architecture, to have reduced number of nodes; this further reduces latency time and improves overall performance. There are three components of this flat structure: A. Evolved Packet Core, which defines network functions, B. User Equipment, i.e. any equipment that contains a Universal Subscriber Identity Module (USIM) for example a smart phone/Tablet/Notebook etc. and C. Evolved Node B(eNB/eNodeB), which is equivalent to a radio base station in GSM/GPRS, but with added functionalities. It controls all radio related functions in the fixed part of the system and works as bridge between UE and the EPC.

eNodeBs are typically distributed throughout the coverage region. The interface between two eNodeBs is known as X2 ;  interface between an eNB and EPC is known as S1 interface. The protocols used between the eNodeBs and the UE are known as the AS protocols. An eNodeB serves multiple UEs at a time but a UE connects to a single eNB at a time. Neighboring eNBs are also connected with each other for handovers. eNBs also contain resource management and logic control functions. This added capability allows eNBs to directly communicate with each other, eliminating the need for mobile switching systems (MSCs) or controllers (BSCs or RNCs).

Other important features of LTE-A include enhanced intercell interference coordination (eICIC), self-optimizing networks (SON) and the deployment of heterogeneous network (HetNet). Hetnet allows various Low Power Nodes (LPNs) to be distributed across a macrocell network as an underlay. The LPNs includes pico eNBs that are typically deployed by the operators or home eNBs which are randomly employed by consumers, relays, and distributed antenna systems (DAS).

LTE-A also supports Device to Device communication (D2D). Thus, UE pair moving within close proximity to each other can establish a D2D link which can be operated either in the unlicensed spectrum band or with the assistance of eNB the link can operate in the same licensed band used by cellular UEs. Furthermore operator assisted D2D communications functions as controlled underlay to the existing LTE-A networks and has added features of transmit power control, peer discovery, physical resource blocks (PRBs) assignment and interference management.

1. **Load Balancing Mechanism**

Load of a cell is measured in terms of usage of different resources with respect to their available limits. These are:

1. Total transmit power
2. Total received power
3. Interference in a cell
4. Cell throughput in downlink/uplink
5. Increase in blocking
6. Handover failure rate

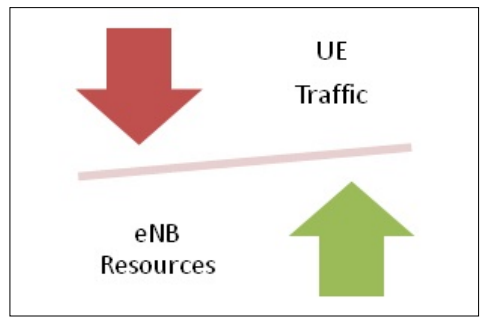

**Figure 2: Load Balancing**

Load balancing techniques may be based on the idle mode users or active mode users.

1.1 Active Mode Load Balancing

Active load balancing allows active mode UEs to be load balanced across cells to lower the overall congestion across cells. The advantage of active load balancing is that the system has a direct measurement mechanism and knowledge of each user's traffic requirements and radio conditions before deciding to load balance. Therefore, with the scheduler and interfaces to other base stations (X2 interface for intra-LTE and/or S1 interface for inter-RAT), it is possible to make accurate decisions for load-based HO. A "load-based HO" reason code is included during handover (HO) messaging to allow the target cell knowledge for admission control.

1.2 Idle Mode Load Balancing

Idle mode load balancing is more difficult to achieve since eNB cannot detect the idle mode users, eNB becomes aware of a user only when it become active or when the tracking area of the user changes and a TAU message is sent by the UE. The way around this immeasurable condition is for the system to adjust cell reselection parameters for the idle users based on the current active user condition, since, in LTE; idle mode inter-frequency load balancing is controlled by the cell reselection procedure. System parameters that control cell reselection and the operator's channel frequency preferences are transmitted to UEs in the System Information Blocks (SIBs). Thus As real-time traffic or QoS demands increase in a cell; it would be possible for the cell to adjust the cell reselection parameters in order to force users nearest the cell edge to select their strongest neighbour or to force a handover to a co-located carrier that has more available resources. Care must be taken to coordinate such parameter adjustments between cells (i.e. utilizing the X2 interface) in order to prevent coverage gaps, as well as to adjust active mode parameters to avoid immediate handover upon an idle to active transition.

Load balancing in LTE is often an autonomous function, and it is modeled as an optimization problem on matching between users and eNBs according to different metrics. One of LB function modules constantly monitors the load of the network; unexpected loading conditions caused by sudden rise in data usage are managed by either adjusting network parameters or the surplus load is shifted from overloaded eNBs to the neighboring eNBs with free resources.

Load balancing optimization comprises of a series of related functions; one of the important functions is the adaptation of handover configuration function enables requesting of a change of handover and/or reselection parameters at target cell, as a part of the load balance procedure. The source cell that initialized the load balancing estimates if the mobility configuration in the source and/or target cell needs to be changed. If the amendment is needed, the source cell initializes mobility negotiation procedure toward the target cell. This is applicable for both idle and active mobility cases. The source cell informs the target cell about the new mobility settings and provides cause for the change such as, load balancing related request. The proposed change is expressed by as the difference (delta) between the current and the new values of the handover trigger. The handover trigger is the cell specific offset that corresponds to the threshold at which a cell initializes the handover preparation procedure. Cell reselection configuration may be amended to reflect changes in the handover setting. The target cell responds to the information from the source cell. The allowed delta range for handover trigger parameter may be carried in the failure response message. The source cell should consider the responses before executing the planned change of its mobility setting. All automatic changes on the HO and/or reselection parameters must be within the range allowed by OAM.

### 1.3 Handover in LTE-Advanced

Handovers are the essential element of load balancing schemes in LTE; thus we will discuss the HO in LTE networks.

#### 3.3.1 Intra-LTE Handover

Intra-LTE handover within the current LTE network is achieved typically by utilizing the X2 interface to exchange load reporting information, to geographically neighboring cells or co-located cells on a different carrier frequency. The load information consists of:

(a) Radio Resource Usage
(b) Uplink/Downlink Guaranteed Bit Rate (GBR)
(c) Physical Resource Block (PRB) usage
(d) Uplink/Downlink non-GBR PRB usage(e) Uplink/Downlink total PRB usage

(f) Hardware (HW) load indicator

(g) Uplink/Downlink HW load: Low, Mid, High, Overload

(h) Transport Network Load (TNL) indicator

(i) Uplink/Downlink TNL load: Low, Mid, High, Overload

(j) Cell Capacity Class value (Optional)

(k) Uplink/Downlink relative capacity indicator

(l) Capacity value

(m) Uplink/Downlink available capacity for load balancing as percentage of total cell capacity

### 3.3.2 Inter-RAT Handover

Inter-RAT HO refers to a load balanced handover between LTE and another network, utilizing the S1 interface to exchange load reporting information. In the event, the handover is to non-3GPP technology, then load reporting on the relevant interfaces still need to be standardized. A dedicated procedure for inter-RAT cell load request / reporting is provided with minimal impact using a generic SON container extension of the RAN Information Management (RIM) mechanism. Load information is provided in a procedure separated from existing active mode mobility procedures, which is used infrequently and with lower priority with respect to the UE dedicated signaling. The load information consists of:

Cell Capacity Class value

Uplink/Downlink relative capacity indicator

Capacity value

Uplink/Downlink available capacity for load balancing as percentage of total cell capacity

Where, Capacity value is expressed in available E-UTRAN resources.

There have been lots of methods for load balancing problem in wireless cellular networks however most of these schemes are designed as per the specifications ( MAC and physical layer) of a particular cellular systems.

First we will discuss few classical approaches for load balancing:.

1) LB using dynamic channel assignment: In this scheme an over loaded cell borrows channels from other less loaded neighbouring cells. This scheme was widely used in circuit switched GSM type systems, where neighbouring cells use different frequency channels. In the case of LTE and LTE-A radio interface is based on OFDMA, where such channel borrowing is not possible due to intra-cell interference; So, this scheme is not applicable in the context of LTE / LTE-A.

2) Coverage Area based Load Balancing: These schemes rely on mechanisms to change the effective coverage area of the cell to in accordance with the load variation. Following techniques can be used to achieve LB :

   1. Pilot power modification
   2. Antenna tilt
   3. Handover (HO) parameter modification

In power adaptation, using pilot power modification scheme the coverage of congested cells is contracted (or expanded) either by reducing (or raising) the pilot power. It has a few drawbacks, for overloaded cells, it decreases the power which may degrade indoor coverage. For low loaded cells more transmission power is required and consequently power amplifiers may be needed.

Second approach is based on smart base station antennas which dynamically changes cellular coverage size and shapes according to load distribution. This approach is limited by the availability of Remote Electrical Tilt (RET) controllers. Apart from these two techniques hybrid approaches involving both antenna tilt and pilot power modification have also been explored however, handover parameter based approach has been most extensively studied in the context of LTE/LTE-A and it is one of the most promising technique because of its flexibility, and effectiveness in complex cellular networks. Handover parameter based load balancing aims at finding the optimum handover (HO) offset value between the overloaded eNB to the target eNB. This scheme is also known as Mobility Load Balancing.

Most of these aforementioned load balancing techniques were designed for cellular networks with users without any quality-of service (QoS) requirements, and therefore only a few may be applicable in the LTE network, which aims to serve users with specific QoS requirements. Further in all the existing solutions the success of LB scheme depends on the availability of low loaded adjacent neighbor cells that can easily take over load from overloaded cells. However, in complex load scenarios observed in LTE Systems adjacent neighbor cells of an overloaded cell may not have enough capacity available. Furthermore, in LTE-A HetNets traffic load has to be balanced among cells having different size, maximum transmit power and other differing performance metrics; consequently, new approaches need to be explored for load balancing in LTE-A networks.

Some of the new approaches being explored for load balancing in LTE-A cellular networks are discussed below:

Dynamic load balancing approach proposed in [8] combines conventional cellular technology and ad hoc wireless networking technology to provide cost efficient solution for heavily loaded networks. A number of ad hoc relay stations (ARSs) are used to relay signals between MHs and base stations; thus using ARSs load is transferred from congested cell to non-congested cell.

Game-Theoretic Approach: A major issue with aforementioned classical load balancing schemes is occurrence of irrational LB, causing a "ping-pong" LB problem, which means that the offloading will be returned to the source cell again in a short period of time. Thus in order to overcome this problem game-theoretic approach [6] for load balancing is being explored by many research groups. It provides an adequate methodology for analysing resource management for self organized wireless cellular networks .further it increases the capacity usage in the network even when the eNBs have different algorithms for load balancing. In PRB game-theoretic approach, each eNB independently decides on the amount of load to maximize its individual utility in an uncoordinated way.

Adaptive neuro-fuzzy Approach: Atyero et al [5] have proposed the use of soft computing approach towards achieving dynamic QoS-aware load balancing. Three key performance indicators namely number of satisfied user, virtual load and fairness distribution index are used to achieve load balancing. Another approach based on D2D communication has been proposed by Jiajia Liu et al [7] which promises to provide efficient traffic management in multi-tier cells according to their real-time traffic distributions.

## 4. Conclusion

The operating conditions in the cellular radio network vary with subscribers' demand which may cause unexpected loading on certain network resources with sudden increase in mobile data usage. Load balancing is performed to offload excess traffic to low loaded similar network elements. Current load balancing optimization processes are handled manually by radio engineers. Self optimized load balancing mechanism is needed for improving efficiency and operation cost reduction. Self optimizing networks increase the efficiency and Quality of service and at the same time reduce the operational cost. Thus the Development of novel self-optimization or self-healing methods will make the planning and management of future generation wireless networks much easier and minimize the needed human involvement.